\documentstyle[prl,aps,graphicx,twocolumn]{revtex}
\begin{document}
\draft
\twocolumn[
\hsize\textwidth\columnwidth\hsize\csname @twocolumnfalse\endcsname
\title{A ratchet driven by quasimonochromatic noise}

\author{M. Array\'as$^{(a)}$, R. Mannella$^{(b,a)}$,
 P. V. E. McClintock$^{(a)}$, A. J. McKane$^{(c)}$, and N. D. Stein$^{(a)}$}

\address{$^{(a)}$Department of Physics, Lancaster University, Lancaster
LA1 4YB, UK\\
$^{(b)}$Dipartimento di Fisica, Universit\`{a} di Pisa and INFM UdR
Pisa,
Piazza Torricelli~2, 56100 Pisa, Italy\\
$^{(c)}$Department of Theoretical Physics, University of Manchester,
Manchester M13 9PL, UK}

\date{\today}
\maketitle
\widetext
\begin{abstract}
The currents generated by noise-induced activation processes in a periodic
potential are investigated analytically, by digital simulation and by
performing analogue experiments. The noise is taken to be quasimonochromatic
and the potential to be a smoothed sawtooth. Two analytic approaches are
studied. The first involves a perturbative expansion in inverse powers of
the frequency characterizing quasimonochromatic noise and the second is a
direct numerical integration of the deterministic differential equations
obtained in the limit of weak noise. These results, together with the digital
and analogue experiments, show that the system does indeed give rise, in
general, to a net transport of particles. All techniques also show that a
current reversal exists for a particular value of the noise parameters.
\end{abstract}
\pacs{PACS numbers: 05.40.Ca, 05.10.Gg, 02.50.Ey}
]
\narrowtext

\section{Introduction}

The nature of microscopic engines, such as molecular motors, has  been the
subject of much research over the last five or six years. This recent activity
was stimulated by the possibility of noise-induced currents
\cite{magnasco,millonas,charlie,buttiker},
and was motivated to a large extent by the
desire to model protein motors. These are proteins which are connected
to a biopolymer and catalyze the conversion of adenosine triphosphate
(ATP) to adenosine diphosphate (ADP).
The energy released by this process is used by the
motor protein to generate motion along the biopolymer in one particular
direction.  This is modelled as a microscopic object moving
unidirectionally along a one-dimensional periodic structure
\cite{reviews}. It is  this problem of rectifying processes at small
scales that has stimulated most  of the theoretical work in this area. A
key ingredient is the presence of random Brownian forces. As a consequence
it is natural to describe these stochastic  ratchets, as they are
frequently called, as a particle moving in a periodic  potential subject
to noise and to formulate this mathematically as a Langevin  equation
\begin{equation}
m\ddot{x} + \alpha \dot{x} + \partial_{x}V(x, t) = \xi(t) \ ,
\label{model}
\end{equation}
where $x(t)$ is the coordinate of the particle, $\alpha$ is a friction
constant, $V(x , t)$ is a periodic asymmetric potential and $\xi(t)$ is the
noise.

Before discussing (\ref{model}) in more detail, we should point out that there
are at least three other reasons for the renewed interest in such systems. The
first is
a new generation of experiments that can be performed in vitro \cite{reviews},
which has led to the increased sophistication of the models now studied.
The second
is the application of these ideas to non-biological situations at small scales
--- the realm of nanotechnology. Finally, this problem throws up fundamental
questions concerning irreversibility and the second law of thermodynamics.
These issues are discussed extensively elsewhere \cite{reviews}, so we will
just make the essential point that it is only if detailed balance holds that we can use
considerations based on the second law to decide that no coherent
unidirectional motion is possible \cite{feynman}.

The classification of different types of ratchets is in fact most easily
carried out with reference to (\ref{model}), since the terminology
used to describe  ratchets (correlation, flashing) is not always
applied consistently. Nearly  all studies neglect the inertial term in
(\ref{model}) and scale time by $\alpha$, so that the coefficient of
the $\dot{x}$ term is unity. Most studies have focussed on the cases
(i) $V(x, t) = V(x)$ is deterministic and $\xi (t)$ is non-white (so
that detailed balance does not hold), (ii) $V(x, t) = V(x)\phi(t)$
where $\phi(t)$ may be deterministic or random \cite{bier} and
$\xi(t)$ is white noise. Since the  main prerequisite for any ratchet
is that the system does not obey detailed  balance and, since detailed
balance may be violated in many different ways, it is clear that many
other forms are possible.

In this paper we will study a ratchet of the type (i) discussed above,
sometimes called a correlation ratchet. In section II the model is described 
in more detail: the potential is specified as is the type of noise ---
quasimonochromatic noise (QMN)~\cite{Dykman:93a}, whose application to
ratchets was first discussed in~\cite{millonas}. We study the model
using analytic techniques in sections III and IV and by the digital and 
analogue simulations which are discussed in section V. We end with an 
analysis of results and conclusions. Some of our preliminary work has 
already been reported \cite{arrayas}.

\section{The model}
In this section we will write down an explicit representation for a
correlation ratchet acted upon by quasimonochromatic noise. Having said
this, it only remains to specify the potential $V(x)$. We have already
indicated that this function should be periodic and asymmetric. A
natural choice would therefore be a sawtooth potential: as shown in
\cite{millonas}, a sawtooth potential is the one which maximises the current.
However, in order to carry out analogue experiments we take the first few 
Fourier modes of such a potential leading to  the form
\begin{equation}
V(x) = 2\cos x + \sin x + \frac{1}{3}\cos(2x)+\frac{1}{10}\cos(3x)\ .
\label{V}
\end{equation}
The sawtooth and the approximation to it, (\ref{V}), are both shown in Figure
1, where we can see that we are modelling the sawtooth potential by rounding
the corners. This not only makes it easier to reproduce in an
analogue  experiment, but it also removes possible singularities in the
theoretical treatment: the \lq\lq sharp corners" at the top and bottom
of the potential would mean that the force that the particle would feel
would be not well defined.

Therefore the model is defined by the Langevin equation
\begin{equation}
\label{lang}
\dot{x} + V'(x) = \xi (t) \ ,
\end{equation}
where $V(x)$ is given by (\ref{V}) and the noise $\xi (t)$ is taken to be
Gaussian with zero mean and correlation function
\begin{equation}
\langle \xi(\omega) \xi(\omega ') \rangle = 2D \, 2\pi \, C(\omega)
\delta (\omega + \omega ') \ .
\label{corr}
\end{equation}
We choose $\xi$ to be quasimonochromatic noise (QMN) since it exactly
suits our purposes: the noise cannot be white, for reasons described in
the last section  and it has the physically appealing feature of having
a peak at a non-zero  frequency in its power spectrum (hence the
name~\cite{Dykman:93a} QMN), while being simple  enough to allow
analytic progress to be
made. Specifically, the noise is  defined by
\begin{equation}
\label{qmn}
C^{-1} (\omega ) = (\omega ^{2} - \omega _{0}^{2} )^{2} +
4\Gamma ^{2} \omega ^{2} \ .
\end{equation}
$C(\omega )$ is sharply peaked at the frequency
$(\omega _{0}^{2} -2\Gamma ^{2} )^{1/2} \approx \omega _{0}$ in the limit
$\Gamma \ll \omega _{0}$, and so we will frequently be working in this regime.
This type of noise can also be viewed as the result of passing white noise
through a harmonic oscillator filter:
\begin{equation}
\label{harm}
\ddot{\xi} + 2\Gamma \dot{\xi} +\omega _{0}^{2} \xi = \eta
\end{equation}
(hence the name ``harmonic''~\cite{muna,lutz},
which is also often used) where the white noise
$\eta$ has strength $D$.

We shall be mounting a three-pronged attack on the problem posed above: an
analytic treatment based a small $D$ approximation, direct digital simulation
of the Langevin equation, to be discussed in section V, and an analogue
simulation also to be discussed in section V. In the case of the analogue
simulation, the first task, before simulating the Langevin equation itself, is
to check the quality of the QMN produced by the analogue circuit. In order to
accomplish this, we examine the QMN spectrum simulated by (\ref{harm}). Two
examples are shown in Figures 2 and 3.

It is evident that there is good agreement between
the noise generated in the simulations and the theoretical result given by
(\ref{qmn}). It is also clear from these figures that the shape of the
spectrum changes considerably depending on the values of the parameters
$\Gamma$ and $\omega$. In fact, these figures illustrate the two regimes for
QMN noise. The first (Figure 2) is an example of the case
$\omega_{0}^{2} > 2\Gamma^{2}$. The spectrum has a local minimum at
$\omega = 0$, rising to a maximum value at
$\omega^{2} = \omega_{0}^{2} - 2\Gamma^{2}$, and then falling off to zero as
$\omega \rightarrow \infty$. As we have remarked already, in the limit
$\omega_{0}^{2} \gg 2\Gamma^{2}$ the peak becomes narrower and better
defined. Moreover, for values of $\omega$ such that $\omega \ll \omega_{0}$,
the spectrum is essentially flat and approximates well white noise. On the
other hand Figure 3 illustrates the case $\omega_{0}^{2} < 2\Gamma^{2}$, where
the spectrum has a local maximum at $\omega = 0$ and   falls away to zero as
$\omega \rightarrow \infty$. So, in summary, if the damping parameter $\Gamma$
is small enough, the power spectrum has a peak at non-zero frequency. As
$\Gamma$ increases, the peak broadens and moves towards zero frequency.  For
$\Gamma$ greater than a critical value of $\omega_0/\sqrt{2}$ the maximum of
the power spectrum is at zero frequency. Our aim is to see how the current 
changes as the noise parameters $\Gamma$ and $\omega_0$ vary.

\medskip

\section{General formalism}
In this section we discuss the approach we will use to explore analytically the
generation of noise-induced currents in the correlation ratchet introduced in
the previous section. The method will involve an asymptotic analysis in the
limit where the noise strength, $D$, tends to zero. To construct the
asymptotic expansion it is first necessary to formulate the problem defined
by the Langevin equation (\ref{lang}) in a different form. There are at least
two different ways to proceed. One is to write down an equivalent
Fokker-Planck equation. Since the noise $\xi(t)$ in (\ref{lang}) is not
white, it is first necessary to convert the process into an equivalent
Markovian one with three degrees of freedom, $(x, \xi, \dot{\xi})$, say.
Thus the Fokker-Planck equation will have the form of time-dependent partial
differential equation in three dimensions. We shall not pursue this method
here, instead we will use the approach of expressing the conditional
probability $\langle \delta ( x - x(t) ) \rangle_{\rm IC}$ as an average over
all possible paths (or realizations of the process) with given initial
conditions, denoted here by \lq\lq IC". These initial conditions specify not
only the initial values of $x(t)$, but also of $\xi(t)$ and of $\dot{\xi}(t)$
at $t=t_0$. The explicit form for the path-integral is \cite{mckaneI}
\begin{eqnarray}
P(x, t | {\rm IC}, t_{0} ) & =  & \langle \delta ( x - x(t) ) \rangle_{\rm IC}
\nonumber \\ \nonumber \\
& = & \int_{\rm IC} \, {\cal D}x {\cal P}[x] \delta ( x - x(t) )\ ,
\label{pathint}
\end{eqnarray}
where ${\cal D}x$ is the appropriate measure defined so that $P$ is correctly
normalized and
\begin{equation}
{\cal P}[x] = J[x]\, \exp{ - S[x]/D}\ .
\label{peeofx}
\end{equation}
Here $S[x]$ is the action functional, which will be discussed in more detail
below, and $J[x]$ is the Jacobian of the transformation from $\eta(t)$ to
$x(t)$, for which we will not require an explicit form.

The method for finding $S[x]$ is discussed in some detail in \cite{mckane96},
but we can obtain it relatively quickly from (\ref{lang})
and (\ref{harm}) by first writing
\begin{eqnarray}
\left[ \dot{x} + V'(x) \right] + \frac{2\Gamma}{\omega_{0}^{2}}\left[ \ddot{x}
+ \dot{x}V''(x) \right] & + & \nonumber \\ \nonumber \\
\frac{1}{\omega_{0}^{2}}\left[ \stackrel{{\bf ...}}{x} + \ddot{x}V''(x)
+ \dot{x}^{2}V'''(x) \right] & = & \omega_{0}^{-2}\, \eta (t)\ .
\label{eta}
\end{eqnarray}
Since the noise $\eta$ is Gaussian, white, with strength $D$,
and has zero mean,
the probability functional $P[\eta]$ has the form
$\exp{ - (1/4D)\int dt\, \eta^{2}(t)}$. A naive substitution of (\ref{eta})
into this expression is sufficient to give the correct functional form for
${\cal P}[x]$ to leading order in $D$, namely
\begin{equation}
{\cal P}[x] = J[x]\, \exp{ - {\cal A}[x]/\Delta}\ ,
\label{peeofx2}
\end{equation}
where
\begin{equation}
{\cal A}[x] = S[x]/\omega_{0}^{4} \ \ \ , \ \
\Delta = D/\omega_{0}^{4}
\label{redef}
\end{equation}
and
\begin{displaymath}
{\cal A}[x] = \frac{1}{4} \int _{-\infty}^{\infty}\!\ dt\, \left\{ \left[
\dot{x} + V'(x) \right] + \frac{2\Gamma}{\omega_{0}^{2}}\left[ \ddot{x}
+ \dot{x}V''(x) \right] \right. +
\end{displaymath}
\begin{equation}
\left. \frac{1}{\omega_{0}^{2}}\left[ \stackrel{{\bf ...}}{x} + \ddot{x}V''(x)
+ \dot{x}^{2}V'''(x) \right] \right\}^{2} \ .
\label{act}
\end{equation}
Having discussed the reformulation of the problem expressed as the Langevin
equation (\ref{lang}) as a Fokker-Planck equation or as a path-integral, we
are now in a position to discuss the $D \rightarrow 0$ asymptotics. In the
case of the Fokker-Planck equation one may perform a WKB-like analysis, while
in the case of the path-integral one may evaluate (\ref{pathint}) by steepest
descents, the paths which dominate the integral being those for which
\begin{equation}
\label{variation}
\frac{\delta {\cal A}[x]}{\delta x} = 0 \ .
\end{equation}
From (\ref{act}) it can be seen that a sixth-order nonlinear differential
equation is obtained. The solutions of this equation, subject to the
appropriate boundary conditions are the instantons or optimal paths,
$x_{c}(t)$, of the model. Substituting this solution back into the action
gives a {\it number} $S \equiv S[x_{c}] = \omega_{0}^{4}{\cal A}[x_{c}]$.
In either case, the WKB treatment or the steepest descent evaluation of the
path-integral, an analysis of the conditional probability (\ref{pathint})
leads to a rate of escape from one potential well to another which has the
characteristic form ${\cal N}\exp{- S/D}$, where ${\cal N}$ will be termed the
prefactor. In the ratchet we are interested in the current $j$ which is
proportional to the difference between the rates of escape from a particular
potential well to the neighboring wells on the right and on the left. It is
therefore reasonable that it should have the form \cite{millonas}
\begin{equation}
j = \lambda\left[ {\cal N}_+ \exp(-S_+ /D) - {\cal N}_- \exp(-S_- /D)\right]\ ,
\label{curr}
\end{equation}
where the plus and minus symbols denote right and leftward transitions
respectively and $\lambda$ is the well spacing.

In the next section we will calculate the actions in (\ref{curr}), by
solving the sixth order equation obtained from (\ref{variation}), numerically.
However, in order to get some intuition for what may happen we will end this
section by assuming that $\omega_{0}$ is large (compared to the scale set by
the curvature of the potential at the bottom of the wells) and obtaining the
action as a power series in $\omega_{0}^{-2}$. In order to do this, we first
rewrite the noise correlator (\ref{qmn}) in the generic form
\begin{equation}
\label{qmn2}
C^{-1} (\omega ) = \omega_{0}^{4} \left[ 1 + \kappa_{1}\tau^{2}\omega^{2}
+ \kappa_{2}\tau^{4}\omega^{4} \right]\ ,
\end{equation}
where
\begin{equation}
\tau = \omega_{0}^{-1}\ , \
\kappa_{1} = -2 \left( 1 - \frac{2\Gamma^{2}}{\omega_{0}^{2}} \right)
\ , \ \kappa_{2} = 1 \ .
\label{kappas}
\end{equation}
With the form (\ref{qmn2}), the action for a path, starting at the bottom of a
well at $x = a$ and ending at the top of an adjacent barrier at $x = b$, is
given by \cite{mckane89}
\begin{eqnarray}
{\cal A} \equiv {\cal A}[x_{c}] & = & \int_{a}^{b} dx \, V'
+ \kappa_{1}\tau^{2} \int_{a}^{b} dx \, V'(V'')^2 \nonumber \\
& + & \kappa_{2}\tau^{4} \int_{a}^{b} dx \,
V'\left[ (V'')^2 + V'V''' \right]^{2} \nonumber \\
& - & \kappa_{1}^{2}\tau^{4} \int_{a}^{b} dx \, (V')^{3}(V''')^2
+ O(\tau^{6}) \ .
\label{genres}
\end{eqnarray}
The $O(\tau^{6})$ terms are also known, they are given in \cite{mckane89}, and
are proportional to $\kappa_{1}^{3}$ and $\kappa_{1}\kappa_{2}$.

First of all, suppose that the spectrum is sharply peaked,
$\omega_{0}^{2} \gg 2\Gamma^{2}$, then both $\kappa_{1}$ and $\kappa_{2}$ have
magnitudes of order one. In this case the expansion (\ref{genres}) is simply
one in powers of $\tau^{2} = \omega_{0}^{-2}$. The $O(\tau^{2})$ corrections
are necessary, otherwise ${\cal A}$ simply depends on the height of the barrier
$\Delta V = \int_{a}^{b}\, dx V'(x)$ and the asymmetry of the potential does
not manifest itself. An interesting special case is when $\omega_{0}$ is large,
but $\omega_{0}^{2} = 2\Gamma^{2}$. Then $\kappa_1$ is zero, and the second and
fourth terms on the right hand side of (\ref{genres}) vanish, as do all of the
$O(\tau^{6})$ terms. So in this case
\begin{eqnarray}
{\cal A} = \Delta V & + & \tau^{4} \int_{a}^{b} dx \,
V'\left[ (V'')^2 + V'V''' \right]^{2} \nonumber \\
& + & O(\tau^{8}) \ \ [\ {\rm for} \ \omega_{0}^{2} = 2\Gamma^{2} \ ] \ .
\label{spcase}
\end{eqnarray}
If there were no $O(\tau^{4})$ terms, it would be the case that when
${\omega_{0}}^{2}=2\Gamma^{2}$, the action only depended on the barrier height,
and so if the height of the barrier was the same to the right or to the left,
we should not expect any net current. Moreover, if we plot the spectrum
of the noise, for the particular value of ${\omega_{0}}^{2}=2\Gamma^{2}$,
we can see that it is very flat: the particle effectively is feeling a white
noise which gives no ratchet effect at all. Since there are $O(\tau^{4})$
corrections in (\ref{spcase}), this is not quite so however. In order to
investigate this point in a little more detail, we have calculated the
integrals in
(\ref{genres}) (and (\ref{spcase})) using $V(x)$ given in (\ref{V}). We find
\begin{equation}
{\cal A}_{\pm}  =  a_{\pm}^{(1)} + a_{\pm}^{(2)}\kappa_{1}\tau^{2}
 +  a_{\pm}^{(3)}\kappa_{2}\tau^{4}
 -  a_{\pm}^{(4)}\kappa_{1}^{2}\tau^{4} + O(\tau^{6}) \ ,
\label{Apm}
\end{equation}
where the coefficients $a_{\pm}^{(i)}$ are given below:

\begin{center}

\begin{tabular}{|c||c|c|c|c|}
\hline
& & & & \\
& \ \ a$^{(1)}$ \ \ & \ \ a$^{(2)}$ \ \ & \ \ a$^{(3)}$ \ \ & \ \ a$^{(4)}$ \
\ \\
& & & & \\
\hline
\hline
& & & & \\
$\ \ + \ \ $ & 4.62 & 22.58 & 681.41 & 751.48 \\
& & & & \\
\hline
& & & & \\
$ - $ & 4.62 & 5.61 & 59.06 & 70.15 \\
& & & & \\
\hline
\end{tabular}

\end{center}

\centerline{Table 1. Numerical values of the integrals}

\centerline{in (\ref{genres}) for the potential (\ref{V}).}

\smallskip

Let us focus on the particular values $\omega_0=10.5$ and $\omega_0=31.6$
which we will use later. A short calculation using (\ref{Apm}) and the values
of $a^{(i)}_{\pm}$ given in the table above shows, that to $O(\tau^{6})$,
${\cal A}_{+}={\cal A}_{-}$ when $\Gamma=6.84$ (for $\omega_0=10.5$) and
$\Gamma=22.13$ (for $\omega_0=31.6$). If we had used the result
${\omega_{0}}^{2}=2\Gamma^{2}$ --- valid for small $\tau$ as indicated by
(\ref{spcase}) --- the corresponding values of $\Gamma$ would have been 7.42
and 22.34 respectively. So we see that $\Gamma = \omega_{0}/\sqrt{2}$ is a
reasonable estimate for the value at which ${\cal A}_{+}$ and ${\cal A}_{-}$
become equal when $\omega_0$ has the larger value (31.6), but it is
considerably different in the case when $\omega_0$ is smaller: $\omega_0=10.5$. We
shall discuss the interpretation of the point where the actions for the left-
and right-moving transitions become equal in more detail in the concluding
section.

\section{Calculation of leading contribution}
In this section we calculate the leading small $D$ contributions $S_{\pm}$ (or
alternatively the leading small $\Delta$ contributions ${\cal A}_{\pm}$) to
the current $j$ in (\ref{current}) for the case of the potential (\ref{V}). In
the preceding section we illustrated the general idea by evaluating these actions for
large $\omega_{0}$, but a general analytic treatment is not possible and we
will have to resort to a numerical calculation of their values.
>From (\ref{act}), the general form of the action functional is
\begin{equation}
{\cal A}[x] = \int^{t}_{t_0} dt \, L( \dot{x}, \ddot{x},
\stackrel{{\bf ...}}{x} ; t)\ .
\label{genact}
\end{equation}
The variation (\ref{variation}) leads to an Euler-Lagrange equation of sixth
order
\begin{equation}
\sum_{j=0}^{3}(-1)^{j}\frac{d^{j}}{dt^{j}}\left(\frac{\partial{L}}
{\partial{x^{(j)}}} \right) = 0 \ ,
\label{EL}
\end{equation}
where $x^{(j)} \equiv d^{j} x/dt^{j}$. A numerical solution will involve the
decomposition of this equation into six first-order differential equations.
A systematic procedure for achieving this is provided by the Hamiltonian
formulation for the generalized mechanics given by (\ref{genact}) and
(\ref{EL}) \cite{mckane95}.

If we carry out this procedure starting from the Lagrangian given by
\begin{eqnarray}
L(x,\dot{x}, \ddot{x}, \stackrel{{\bf ...}}{x} ) & = &\frac{1}{4}\{[ \dot{x}
+ V'(x)] +\frac{2\Gamma}{\omega_0^{2}}[\ddot{x}+\dot{x}V''(x)]
+ \nonumber\\
& &\frac{1}{\omega_0^{2}}[ \stackrel{{\bf ...}}{x} + \ddot{x}V''(x) +
\dot{x}^{2}V'''(x)]\}^{2}\ ,
\label{lagrangian}
\end{eqnarray}
we find the following Hamiltonian
\begin{eqnarray}
H(\vec{x},\vec{p}) &=& p_1x_2 + p_2x_3 + \omega_{0}^{4}{p_3}^{2} -
p_3\{ \omega_{0}^{2}(x_2+V')+ \nonumber\\
& &2\Gamma(x_3+x_2V'')+x_3V''+{x_2}^{2}V'''\}\ ,
\label{ham}
\end{eqnarray}
where $\vec{x} = (x_{1}, x_{2}, x_{3})$ and $\vec{p} = (p_{1}, p_{2}, p_{3})$.
The action turns out to be \cite{mckane95}
\begin{equation}
{\cal A} = \int_{-\infty}^{\infty} \!\ p_3^2 \, dt \ .
\label{act2}
\end{equation}
\noindent
Hamilton's equations have their usual form:
\begin{equation}
\dot{x}_i = \frac{\partial H}{\partial p_i} \;\;\;, \;\;\;
\dot{p}_i=-\frac{\partial H}{\partial x_i} \;\;\;, \;\;\; i=1 \ldots 3 \ ;
\label{he}
\end{equation}
they are, by construction, equivalent to the Euler-Lagrange equations
(\ref{EL}). Using the Hamiltonian (\ref{ham}), the six equations (\ref{he})
yield
\begin{eqnarray}
\dot{x}_1 &=& x_2 \label{syst1}\\
\dot{x}_2 &=& x_3 \\
\dot{x}_3 &=& 2\omega_{0}^{4} p_3 - \{\omega_{0}^{2}[x_2+V'(x_1)]
 +2\Gamma[x_3+x_2V''(x_1)]\nonumber \\
& &+x_3V''(x_1)+{x_2}^{2}V'''(x_1)\} \\
\dot{p}_1 &=& p_3\{ \omega_{0}^{2}V''(x_1)
 +2\Gamma x_2V'''(x_1))\nonumber \\
& &+x_3V'''(x_1)+{x_2}^{2}V''''(x_1)\}\\
\dot{p}_2 &=&-p_1 + p_3 \{ \omega_{0}^{2} + 2\Gamma V''(x_1) +
2x_2V'''(x_1) \}\\
\dot{p}_3 &=& -p_2 + p_3 \{ 2\Gamma + V''(x_1)\}\ .
\label{syst2}
\end{eqnarray}
For an escape problem, we are searching
for solutions that provide the minimum of the action. Imposing the
condition that the variation with time of the action is zero, as
before, $H =0$, and following previous work \cite{mckane95}, we choose
as boundary conditions for the ``uphill'' solution (going from the
bottom to the top of the potential)
\begin{eqnarray}
x_1(-\infty)=x_{\rm min} &,& x_1(\infty)= x_{\rm max} \nonumber\\
x_2(\pm \infty) &=& 0 \nonumber \\
x_3(\pm \infty) &=& 0 \ .
\label{bc}
\end{eqnarray}
In order to solve (\ref{syst1})-(\ref{syst2}) in practice we have to
truncate them to a large, but finite, time interval and use the boundary
conditions
\begin{eqnarray}
x_1(-T)=x_{\rm min} &,& x_1(0)= x_{\rm max} \nonumber\\
x_2(-T) = 0 &,& x_2(0) = 0\nonumber \\
x_3(-T) = 0 &,& x_3(0) = 0\ ,
\label{finbc}
\end{eqnarray}
where we have used the time translation invariance of the equations.

The TWPBVP subroutines developed for solving two-point boundary value
problem were used \cite{cas91}. In order to get convergence we used the
following procedures. We linearized the equations
(\ref{syst1})-(\ref{syst2})
at the initial point, and changed the boundary conditions at that
point, perturbing them in the direction of the unstable manifold given by
eigenvectors with eigenvalues having positive real part. We took as an initial
guess a straight line joining the boundary points, and the
solution of this modified problem was used as initial guess to the
original problem, leading to a solution for the optimal path.

Having found this solution, the action
\begin{equation}
{\cal A} = \int_{-T}^{0} \!\ p_3^2 \, dt
\label{act3}
\end{equation}
was calculated, and to minimize the effect of the cut-off effect we added the
correction from integrating the linear expressions on the boundaries from
$-\infty$ to $-T$ and from $0$ to $\infty$. When $T$ is chosen to be large
enough, the result is independent of its value.

The results of this calculation will be discussed in section VI where they will
be compared to the simulations that we have also carried out on this problem.
We now turn to a discussion of these simulations.

\section{Analogue and Numerical Simulations}
In this section we study our correlation ratchet using analogue and
digital simulations. We have measured the variation of the current with
$\Gamma$ in two ways: analogue simulation experiments on an
electronic circuit and Monte-Carlo simulations on a digital computer.

The first of these techniques
\cite{Fronzoni:89,Mcclintock:89,Luchinsky:98a} involves the building of
an electronic circuit to model the system under study, the application
of appropriate forces, and analysis of the response by means of a
digital computer. The absence of truncation errors makes analogue
simulations especially valuable for use e.g.\ with fast oscillating
systems where the integration time (the time over which data are
accumulated and perhaps ensemble-averaged) substantially exceeds the
vibration period, as occurs with QMN.  Digital techniques have the
advantage that they can always {\it in principle} be made more accurate
than analogue methods, which typically achieve 2--3\% accuracy, but the
relative simplicity of analogue simulations and  their realism (being
much closer to a real experiment than a digital simulation) represent
significant advantages.

The electronic circuit used to model (\ref{lang}) and (\ref{harm}) is shown
as a block diagram in Figure 4. The lower section is the harmonic
oscillator used as a ``filter'' to convert \cite{Dykman:93a} quasiwhite
noise from a feedback shift-register noise-generator
\cite{Faetti:84,Luchinsky:98a} into QMN. The QMN is then applied to the
input of the upper part of the circuit, which models the ratchet
potential itself. Although the basis of the circuit is standard
\cite{Luchinsky:98a}, several points of detail deserve amplification.
The force corresponding to the trigonometric potential of (\ref{V}) is
created using trigonometric identities to write it as:
\begin{equation}
\label{trigforce}
V'(x) = -\frac{29}{10}\sin x + \cos x - \frac{4}{3}\sin x \cos x
        +\frac{6}{5}\sin^{3} x\ ,
\end{equation}
\noindent
so we can build the force using only two AD639 ICs \cite{ad}. An
inherent limitation of the AD639 IC is that it can only treat a
restricted range of angles ($\pm 500^\circ$). To prevent its input from
straying outside this range, provision is made for resetting the circuit
using DG303AC \cite{slx}
switches (they have not been plotted in the block diagram)
\cite{Fronzoni:89,Mcclintock:89,Luchinsky:98a}.
The voltage in the circuit corresponding to coordinate $x$ was
digitized with a 12-bit Microstar ADC \cite{Microstar}, model DAP
3200a/415. Data analysis exploited the on-board 100 MHz Intel 486DX
co-processor, which was operated within a MatLab-based PC software
system developed by Kaufman \cite{Kaufman:99}.

The digital simulations were done using a specialized algorithm, described
in \cite{Dykman:93a}, which we will briefly recall here. The particular
structure of (\ref{lang}) and (\ref{harm}) poses, in principle, a problem
if a simple minded algorithm is used in the integration: (\ref{harm})
is characterised by two time scales ($\omega_0$ and $\Gamma$) and the
integration time step (call it $h$) used in the digital simulations would
be chosen in such a way that both $\omega_0 h \ll 1 $ and $\Gamma h \ll 1$.
Also, if $\tau_r$ is the typical relaxation time in (\ref{lang}), we
should also satisfy $h/\tau_r \ll 1$. This latter inequality, for the
typical parameters which are of physical interest, is normally satisfied as
soon as the former one is: in other words, in a simple minded algorithm the
constraint on the integration time step comes from (\ref{harm}) rather
than from (\ref{lang}), because the time scales involved by $\omega_0$
and $\Gamma$ are (much) smaller than $\tau_r$. From the point of view of
the problem we are trying to solve, however, this would not be very
efficient: we would be using most of the CPU time integrating the noise
equation (\ref{harm}) rather than integrating the dynamical equation
representing the model under study.

The particular structure of (\ref{lang}) and (\ref{harm}) suggests that a
specialized algorithm could be more efficient: the point is that
(\ref{harm}) is a linear filtering of an uncorrelated Gaussian noise.
This means that the output of this equation (the variable $\xi$) is itself a
Gaussian variable, of unknown intensity and correlation: hence, it may be
possible to integrate (\ref{lang}) directly, working out the appropriate
integration algorithm, using the statistical properties of the Gaussian
variable $\xi$. The algorithm used to integrate (\ref{lang}) is the Heun
algorithm~\cite{Greiner:88} which prescribes that we
integrate (\ref{lang}) with a couple of elementary steps, namely we
first predict
\begin{equation}
\tilde x (h) = x(0) + h[-V'(x(0))] + r(h)\ ,
\end{equation}
\noindent
and then correct as
\begin{equation}
x(h) = \{ \tilde x (h) + x(0) + h [-V'(\tilde x(h))] + r(h)\}/2\ .
\end{equation}
In the expression above, we need to evaluate the quantity $r(h) \equiv
\int_0^h \xi(s) \; ds$, which can be written~\cite{Dykman:93a} as
\begin{equation}
r(h) = A_{31} \xi(0) + A_{32} \dot{\xi} (0) + w_3\ ,
\end{equation}
\noindent
where
\begin{equation}
w_3 = B_{31} z_1 + B_{32} z_2 + B_{33} z_3\ ,
\end{equation}
\noindent
and where $z_i$ are uncorrelated Gaussian deviates of average zero and
standard deviation one. Note that the quantity $r(h)$ turns out to be a
linear combination of Gaussian variables, as expected. Defining $\Omega^2
\equiv \omega_0^2 - \Gamma^2$ and $\lambda_\pm \equiv -\Gamma \pm
\sqrt{\Gamma^2 - \omega_0^2}$, we have
\begin{eqnarray}
A_{31} & = &\frac{i}{2\Omega}
\left\{ \frac{\lambda_{-}}{\lambda_{+}} (e^{h \lambda_{+}}-1) -
\frac{\lambda_{+}}{\lambda_{-}} (e^{h \lambda_{-}}-1) \right\} \\
A_{32} & = &\frac{i}{2\Omega} \left\{ \frac{e^{h\lambda_{-}}-1}{\lambda_{-}}-
\frac{e^{h \lambda_{+}}-1}{\lambda_{+}}\right\}\ .
\end{eqnarray}
The expressions for $B_{ij}$ are very cumbersome, and we refer the reader
to~\cite{Dykman:93a}: note that ~\cite{Dykman:93a} contains a misprint, the
quantity $4 \pi T/2 \Omega^2$ on the right hand side of Eq.~(A14) should read
$\Gamma T/\Omega^2$.

A warning is in order concerning the random noise generator. The noise
intensities of interest are fairly small compared to the barrier that the
Brownian particle has to overcome to diffuse and generate a net current. It is
then of great importance to make sure that the rare activation events are
correctly generated, which implies that the noise generator should be
particularly accurate in generating the tails of the distribution. The generator
used works by generating a Gaussian random variable, using the
Ziggurath algorithm~\cite{Marsaglia:84}, from flat random distributions
obtained with a subtract and carry algorithm~\cite{Luscher:94,James:94}.

The actions $A_{\pm}$ were calculated from the slope of
plotting the logarithm of the mean escape time, calculated as the total time of
observation divided by the number of transitions to the left or right,
versus $1/D$.
The current in the experiments is easily obtained by keeping track of the
distance moved by the random walker and dividing it by the total
simulation time. These results can be observed in the figures.

\section{Analysis of Results and Conclusions}
In this section we wish to compare the theoretical predictions of sections III,
IV with the experiments. Our main aim is to understand the structure of the
current (\ref{curr}). This is made up of actions $S_{\pm}$ and prefactors
$\cal{N}_{\pm}$. For small $D$, the action dominates, so we begin by comparing
the actions calculated from (\ref{genres}) with the numerical method discussed
in section IV. For a typical value of $\omega_0=10.5$, the results are shown
in Figure 5. This shows reasonable agreement between analytical and numerical
results of solving the full set of equations of section IV. This fully
justifies the approximation of section III, which is very useful given the
difficulty of carrying out the numerical integrations in the method of section
IV.

Now we are in a position to compare the theoretical predictions with the
digital simulation of section V. This is shown in Figure 6. The digital
simulation shows the same trends as the analytical results, but there is a
large amount of scatter. Nevertheless, the value of $\Gamma$ at the point
where $\cal{A}_{+}=\cal{A}_{-}$ predicted by the digital simulations is in
reasonable agreement with the theoretical value.

The prefactors are, unfortunately, difficult to calculate. In fact, a
calculation for QMN has not yet been carried out. The prefactors are, however,
known for white noise and exponentially correlated noise (for small noise
correlation time \cite{mckaneIII}), and we use these in the expectation that
they are a reasonable approximation to the true result. In Figure 7, the
current calculated from (\ref{curr}) using these prefactors together with an
action calculated as in section IV, is plotted with the data from the digital
simulation. A possible interpretation of the deviation of the theoretical 
from the experimental results could be the approximate prefactor. However it 
is clear that the exponentially correlated prefactor is an improvement over 
the white noise one, and this suggests that the correct QMN prefactor might 
give even better agreement. In any case, it is clearly demonstrated that a 
ratchet consisting of an asymmetric periodic potential plus
quasimonochromatic noise forcing can indeed give rise to a net transport of
particles.

Table 2 displays the current obtained for $\omega_0=10.5$ in both the analogue
experiment and the digital simulation. In this case, although actions can
be calculated analytically, the parameter in the expansion for the
prefactor is no longer small and one is unable to obtain an analytic expression
for the current. The values $\omega_0=10.5$ and $\omega_0=31.6$ were chosen
for technical reasons connected with carrying out the analogue experiment.

\begin{center}

\begin{tabular}{|c|c|c|c|}
\hline
  $\Gamma$ & $\Delta$ & $j_A$ & $j_D$ \\
\hline
  0.938 &  0.564 & $  2.17\times 10^{-4}$ & $  1.53\times 10^{-3}$\\
  1.346 &  0.591 & $  2.31\times 10^{-4}$ & $  1.26\times 10^{-3}$\\
  2.386 &  0.617 & $  2.14\times 10^{-4}$ & $  8.10\times 10^{-4}$\\
  5.250 &  0.654 & $  1.20\times 10^{-4}$ & $  1.74\times 10^{-4}$\\
  6.402 &  0.670 & $  4.32\times 10^{-5}$ & $  1.03\times 10^{-5}$\\
  7.721 &  0.673 & $ -1.29\times 10^{-5}$ & $ -1.78\times 10^{-4}$\\
  9.375 &  0.681 & $ -1.25\times 10^{-5}$ & $ -3.15\times 10^{-4}$\\
 11.170 &  0.693 & $ -3.84\times 10^{-5}$ & $ -4.64\times 10^{-4}$\\
 13.462 &  0.713 & $ -1.20\times 10^{-4}$ & $ -6.54\times 10^{-4}$\\
 17.500 &  0.737 & $ -2.13\times 10^{-4}$ & $ -7.78\times 10^{-4}$\\
 21.000 &  0.796 & $ -3.60\times 10^{-4}$ & $ -1.12\times 10^{-3}$\\
 26.250 &  0.832 & $ -4.33\times 10^{-4}$ & $ -1.28\times 10^{-3}$\\
 29.167 &  0.863 & $ -5.44\times 10^{-4}$ & $ -1.42\times 10^{-3}$\\
\hline
\end{tabular}

\end{center}

\centerline{Table 2. Analogue ($j_A$) and digital ($j_D$)}

\centerline{currents for $\omega_0=10.5$}.

\smallskip

There is no reason to assume that the current reversal ($j = 0$) necessarily 
occurs when ${\cal A}_{+} = {\cal A}_{-}$, because the prefactors may cause 
some deviation from this leading order result. However, from Figure 7 and 
Table 2 it seems that they do in fact occur at the same point --- even though
the magnitude of the analogue current is consistently less than that of the
digital current. This may be because, as $D \rightarrow 0$, the action 
completely dominates or because the prefactors happen to be approximately 
equal at this point.

In the literature, this problem has been already discussed in~\cite{millonas}
and~\cite{bart}. In~\cite{millonas} it was found that there is indeed a net
current in the system, and, working in the limit of small $\Gamma/\omega_0$, 
the authors were able to show that the sign of the current changes as the 
curvature at $\omega=0$ (i.e. $C''(0)$) changes with varying $\Gamma$. 
In~\cite{bart} the authors considered a model of the form~(\ref{lang}), but 
with an additional white noise. However, as we have seen, there is no need to 
introduce an additional noise of this type in order to see a current reversal.
The authors found that the current should change sign for at least two 
different values of $\Gamma$.

Our aim has been to study how the current changed as the noise parameters 
$\Gamma$ and $\omega_0$ varied, for finite $\Gamma/\omega_0$. We found that
that the change in the spectral density curvature at $\omega=0$ mentioned
above is still the main effect in determining the current direction, in 
agreement with~\cite{millonas}. We have included in the theoretical treatment 
higher order terms in $\Gamma/\omega_0$: our result coincides with the result 
of~\cite{millonas} in the appropriate limit, with a small shift in the 
transition point if $\Gamma/\omega_0$ is finite. The simulations which were 
carried out support the theoretical conclusions. We have not observed more 
than one current reversal experimentally (with fixed $\omega_{0}$ and varying 
$\Gamma$), but examination of (\ref{Apm}) shows that there is another solution
(to $O(\tau^{4})$) for which ${\cal A}_{+} = {\cal A}_{-}$. It would be 
interesting to explore this regime in more detail experimentally.

\acknowledgements

We are grateful to M. I.\ Dykman for valuable discussions and
encouragement during the early stages of the project. This work was
supported in part by the Engineering and Physical Sciences Research
Council under grant GR/K49966, and by INTAS  under grant 97-574.

\begin{figure}
\centering
\includegraphics*[angle=-90, width=3.4in]{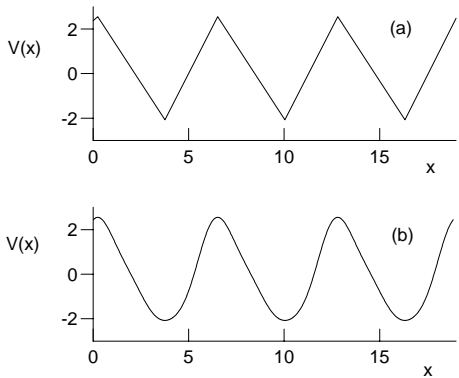}
\caption{\label{sawtooth}(a) Sawtooth potential. (b) `Smoothed sawtooth'
potential.}
\end{figure}

\begin{figure}
\centering
\includegraphics*[angle=-90, width=3.4in]{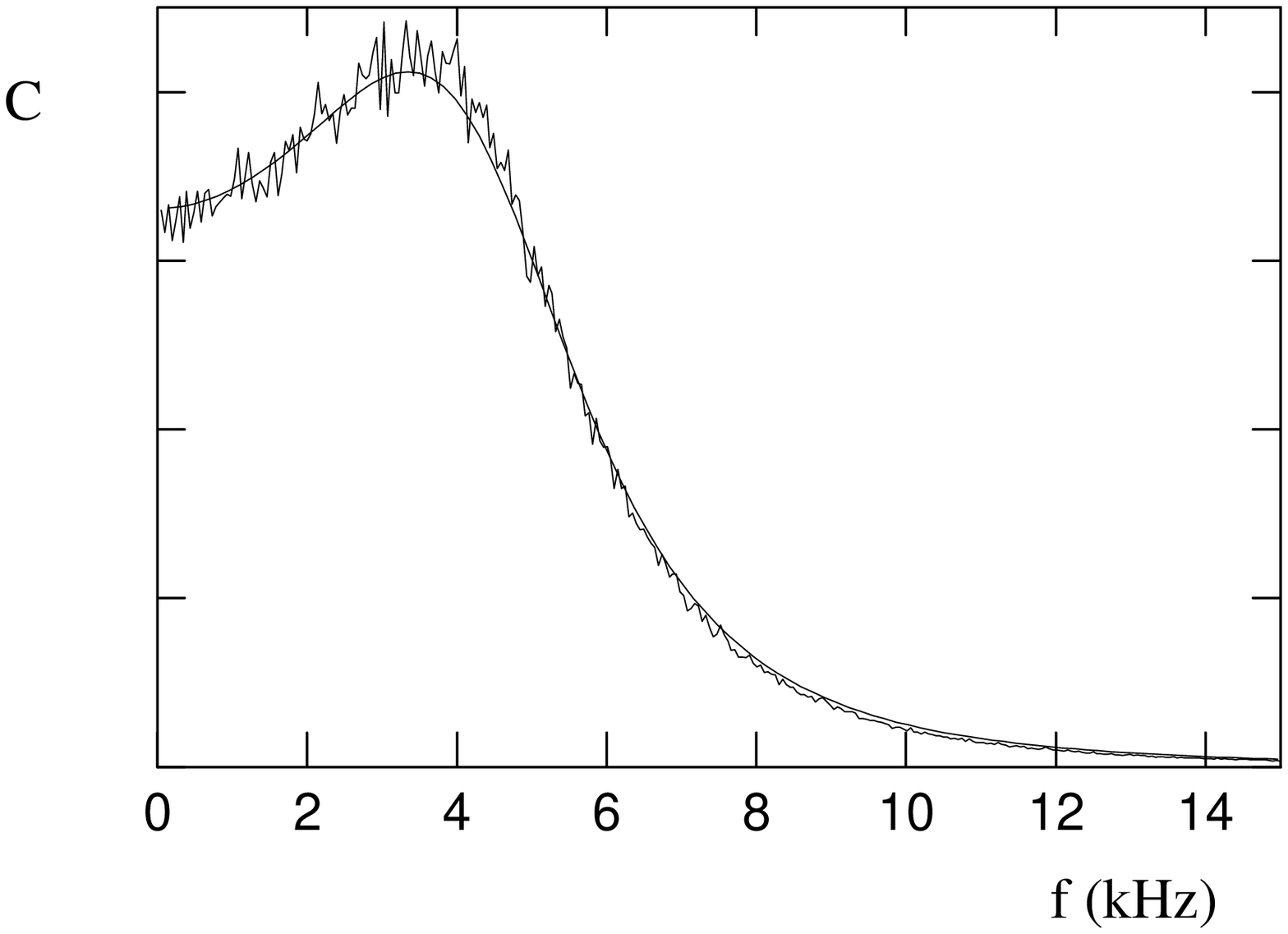}
\caption{\label{powerspec1}Noise power spectrum for $\omega_0 = 31.6$,
$\Gamma = 16.7$. The jagged line is from experiment, and the smooth one
from theory. The frequency $f=\omega/2\pi$.}
\end{figure}

\begin{figure}
\centering
\includegraphics*[angle=-90, width=3.4 in]{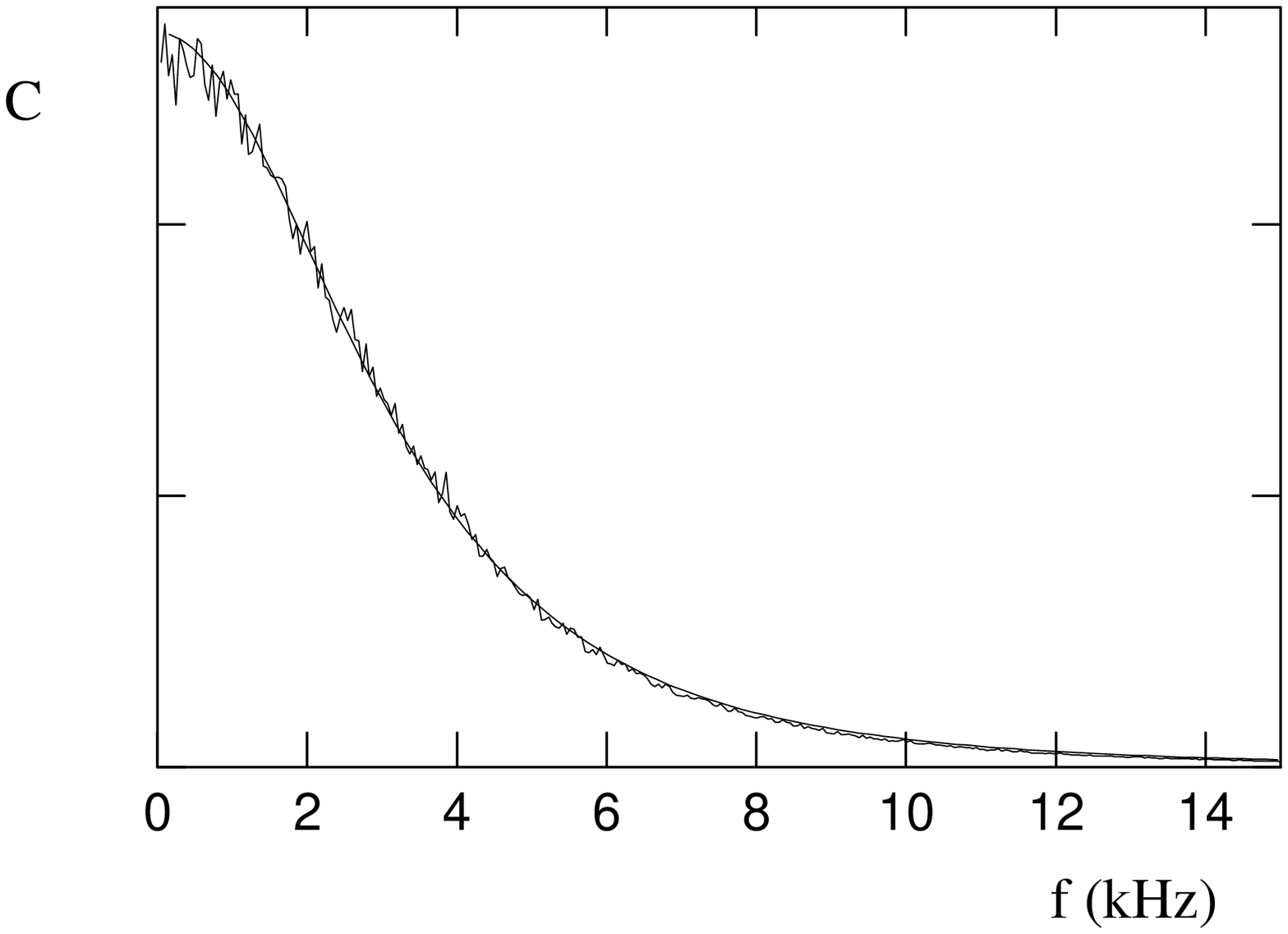}
\caption{\label{powerspec2}Noise power spectrum for $\omega_0 = 31.6$,
$\Gamma = 33.3$.  The jagged line is from experiment, and the smooth one
from theory. The frequency $f=\omega/2\pi$.}
\end{figure}

\begin{figure}
\centering
\includegraphics*[angle=-90, width=3.4 in]{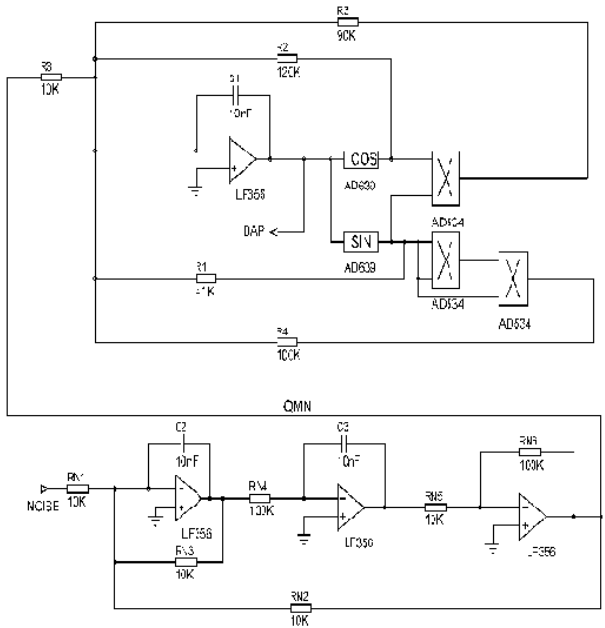}
\caption{\label{circuit}Experimental analogue circuit.}
\end{figure}

\begin{figure}
\centering
\includegraphics*[width=3.1 in]{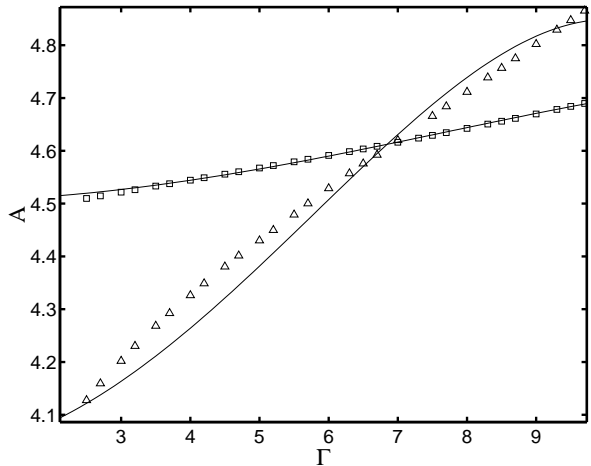}
\caption{\label{action10}Analytic (small $\tau$) versus numerical (general
$\tau$) actions for $\omega_0 = 10.5$. Squares  are for escapes to the left,
triangles to the right. Symbols are from numerical integration, solid lines
from analytic calculations.}
\end{figure}

\begin{figure}
\centering
\includegraphics*[width=3.1 in]{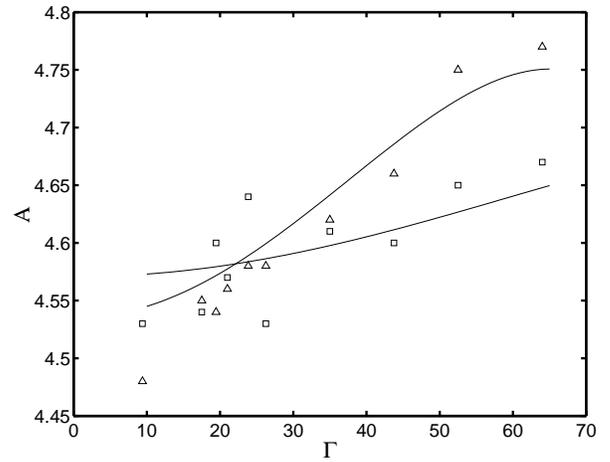}
\caption{\label{action6}Digital simulations and theoretical actions, for
$\omega_0
= 31.6$. Squares denotes escapes to the left, triangles to the right. Symbols
are simulations, curves theory.}
\end{figure}

\begin{figure}
\centering
\includegraphics*[angle=-90, width=3.4 in]{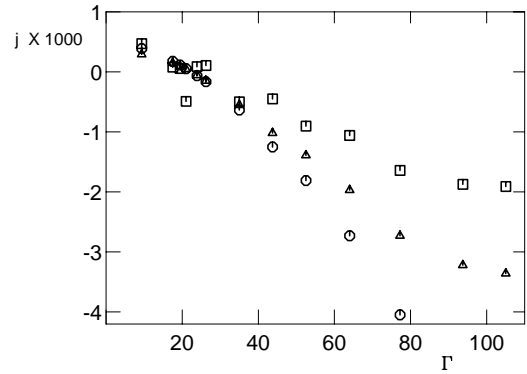}
\caption{\label{current}Current for $\omega_0 = 31.6$ and $\Delta = 
D/\omega_{0}^{4} = 0.87$.
The squares are from digital simulation;
the circles represent theory for a white noise prefactor; the
triangles represent theory with the improved prefactor.}
\end{figure}

\end{document}